\documentclass[a4paper,twoside,10pt]{letter}
\usepackage{graphicx,saj,multicol,subeqnarray,amsmath,url}

\def\papertype{Original Scientific Paper}

\def\p0{\phantom{0}}
\def\it{\sl}

\def\udc{...}
\begin{document}
\baselineskip=3.1truemm
\columnsep=.5truecm
\newenvironment{lefteqnarray}{\arraycolsep=0pt\begin{eqnarray}}
{\end{eqnarray}\protect\aftergroup\ignorespaces}
\newenvironment{lefteqnarray*}{\arraycolsep=0pt\begin{eqnarray*}}
{\end{eqnarray*}\protect\aftergroup\ignorespaces}
\newenvironment{leftsubeqnarray}{\arraycolsep=0pt\begin{subeqnarray}}
{\end{subeqnarray}\protect\aftergroup\ignorespaces}
%


\markboth{\eightrm Solar flares and VLF and LF radio signals}
{\eightrm D. M. {\v S}ULI{\' C} and V. A. SRE{\' C}KOVI{\' C}}

{\ }

\publ

\type

{\ }


\title{A comparative study of measured amplitude and phase perturbations of VLF and LF radio signals induced by solar flares}


\authors{D. M. \v Suli\' c$^1$, V. A. Sre\' ckovi\' c$^2$ }

\vskip 3mm


\address{$^1$University Union - Nikola Tesla Belgrade, Serbia}
\Email{desankasulic}{gmail.com}
\address{$^2$Institute of Physics, University of Belgrade, P.O. Box 57, Belgrade, Serbia}
\Email{vlada}{ipb.ac.rs}


\dates{August 2013}{September 2013}


\summary{Very Low Frequency (VLF) and Low Frequency (LF) signal perturbations were examined to study ionospheric disturbances induced by solar X-ray flares in order to understand the processes involved in propagation VLF/LF radio signals over short paths and to estimate specific characteristics of each short path. The receiver at Belgrade station is constantly monitoring the amplitude and phase of coherent and subionospherically propagating LF signal operated in Sicily, NSC at 45.90, kHz and VLF signal operated in Isola di Tavolara ICV at 20.27 kHz, with great circle distances of 953 km and 976 km, respectively. A significant number of similarities between these short paths is a direct result of both transmitters and receiver geographical location. The main difference is in relation to transmitter frequencies. From July 2008 to February 2014 there were about 200 events that had been chosen for further examination. All selected examples showed that the amplitude and phase on VLF and LF signals were perturbed by solar X-ray flares occurrence. This six-year- period covers both minimum and maximum of solar activity.

Simultaneous measurement of amplitude and phase on the VLF/LF signals during solar flare occurrence was applied for the purpose of evaluating electron density profile versus altitude and as well as for carrying out the function of time over middle Europe.}



\begin{multicols}{2}


\section{1. INTRODUCTION}

The ionosphere is the part of the atmosphere that contains ionized gases. The primary process is photoionization of thermospheric gases by the Sun's extreme ultraviolet radiation, and X-rays. Both of these radiations are $\sim 100$ times stronger at solar maximum than at solar minimum. Secondary processes include ionization by photoelectrons and scattered or reemitted radiation. The ionosphere, at all latitudes, has a tendency to separate in different regions, despite the fact that different processes dominate in different latitudinal domains. The regions: D, E, and F with two layers F$_{1}$ and F$_{2}$ are distinct only in the daytime ionosphere at mid-latitudes. The different regions are generally characterized by density maximum at a certain altitude and density decreases with altitudes on both sides of the maximum.

The lowest region of the ionosphere the D-region, $50 \leq h \leq 90$ km, is formed primarily by the action of solar Lyman-alpha radiation (121.6 nm) on nitric oxide. At night, the ionospheric plasma densities decreases most rapidly at the lowest altitudes and at 90 km altitude it decreases from $10^{11}$ to $10^{8}$ m$^{-3}$ (Schunk and Nagy, 2000).

Very Low Frequency (VLF, 3 - 30 kHz), and Low Frequency (LF, 30 - 300 kHz) radio signals propagate inside the waveguide formed by the lower ionosphere and the Earth's surface (Wait and Spies,1964; Mitra, 1974). A range of dynamic phenomena occurs in D-region, and some of them are: \emph{diurnal effect (day/night), a seasonal effect (summer/winter), correlation with solar activity (sunspot level and solar flares), effects of lightning induced electron precipitation and red sprites}. All these phenomena are followed by changes in electron density of lower ionosphere, which affects the subionospheric VLF/LF propagation as an anomaly in amplitude and/or phase.

During solar flare X-ray irradiances rapidly increase and X-rays with wavelengths below 1 nm are able to penetrate to the D-region, causing ionization of the neutral constituents, predominantly nitrogen and oxygen (Mitra, 1974).

 \section{2. EXPERIMENTAL SETUP}

The perturbations in the D-region induced by solar flares, the Sudden Ionospheric Disturbances, (SIDs), were studied using monitored amplitude and phase data from VLF/LF transmitters, in period July 2008 - February 2014. This period of six years includes minimum and maximum of solar activity. All data were recorded at the Belgrade station (44.85$^{0}$ N, 20.38$^{0}$ E) by the Stanford University ELF/VLF Receiver Atmospheric Weather Electromagnetic System for Observation Modeling and Education (AWESOME). Such data are digitize and generally saved in two different resolutions - high resolution (50 Hz) and low resolution (1 Hz). Narrowband data can be recorded in a continuous fashion, even in case when as many as 15 transmitters are being monitored.

One of the best defined signals received by AWESOME system at Belgrade station originates at the NSC transmitter from Sicily Italy at 45.90 kHz. Other signals received at Belgrade station include:  DHO (Germany 23.4 kHz), HWU (France 18.3 kHz), GQD (UK 22.21 kHz), ICV (Italy 20.27 kHz), NRK (Iceland 37.5 kHz), NAA (USA 24 kHz) and NWC (Australia 19.8 kHz). Great circle distances for those signals are in the range from 1 Mm to 12 Mm.

This paper presents the results of short-distance (propagation distance less than 1000 km) subionospherically VLF/LF propagation for detection SIDs in the D-region. Locations of the transmitters and receiving site are presented in Figure 1. The receiver keeps monitoring the amplitude and phase of coherent and subionospherically propagating VLF signals operated in Sicily (38.00$^{0}$ N, 13.50$^{0}$ E, NSC at 45.90, kHz) and in Isola di Tavolara (40.88$^{0}$ N, 9.68$^{0}$ E, ICV at 20.27 kHz), both in Italy, with great circle distances of 953 km and 976 km, respectively. Propagation paths are southwest - northeast oriented. Both signals propagate over sea, ground, sea and ground, which implies very similar conductivity properties of the waveguide bottoms.

 \vskip.5cm
\centerline{\includegraphics[width=\columnwidth,
height=0.75\columnwidth]{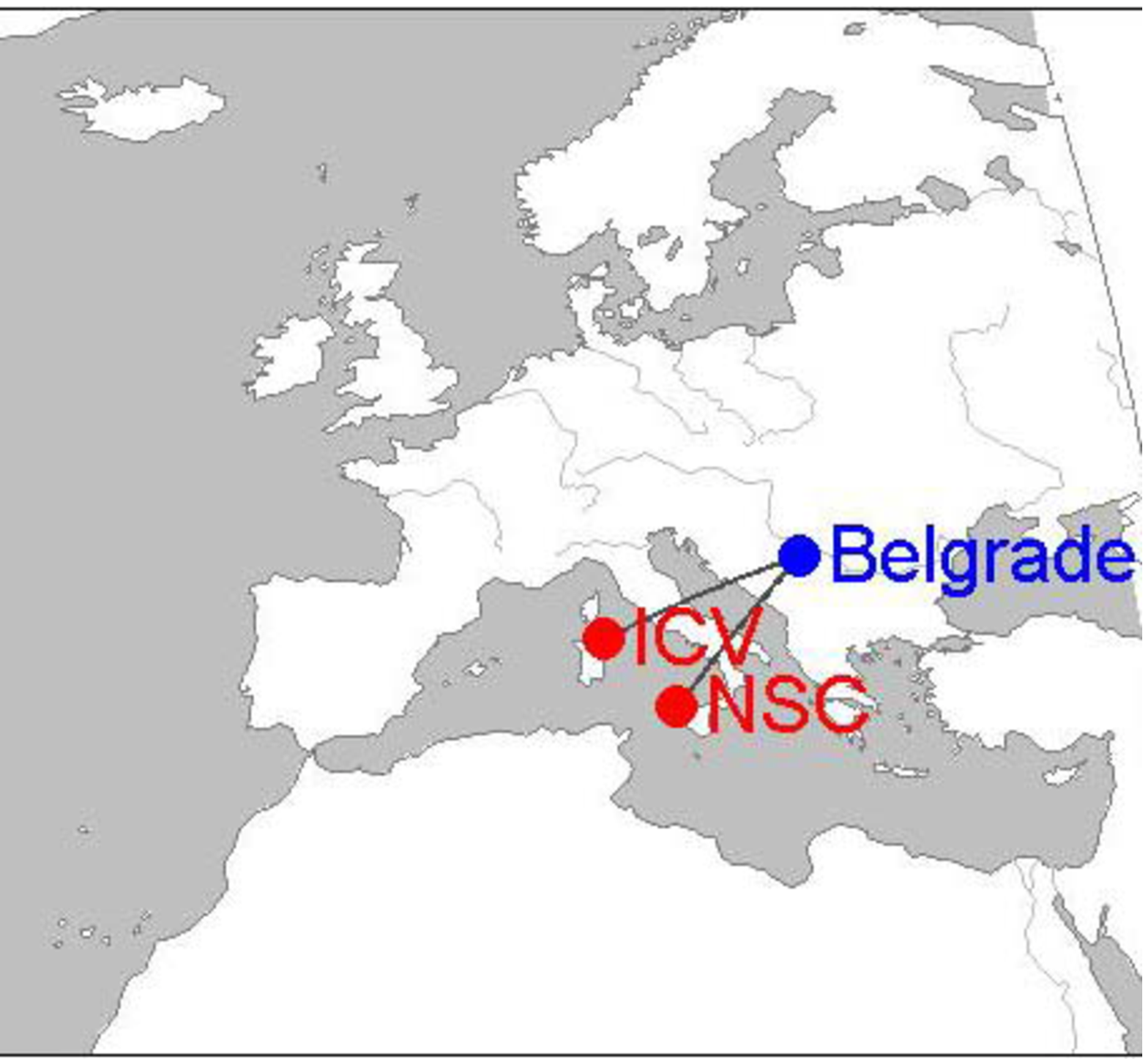}}
\figurecaption{1.}{Great Circle Paths (GCP) of subionospherically propagating VLF signals recorded at Belgrade station. The receiver is continuously monitoring the amplitude and phase of coherent and subionospherically propagating VLF signals operated in Sicily at 45.90, kHz and Isola di Tavolara at 20.27 kHz}

For studying solar flare events in similar conditions we have examined only those events that occurred in the time sector for zenith angle $\chi \leq 60^{0}$. During local winter and equinox this time sector is between 4 or 5 hours around local noon. Local noon is at $\sim$ 10:50 UT. During local summer the time sector lasts about 7 hours. As for the above mentioned facts, we have already examined around 200 solar flare events and analyzed their effects on propagation characteristics of VLF/LF signals.

This work deals with a typical X-ray irradiance $I_{X}$[Wm$^{-2}$] recorded by GOES - 15 satellite for wavelengths ranging from 0.1 to 0.8 nm, available from USA National Oceanic and Atmospheric Administration (NOAA) via the web site: www.swpc.noaa.gov/ftpmenu/lists/xray.html.

\section{3. MEASURED DATA }

The time stability of NSC transmitter proved to be continuous day and night monitoring not only of the amplitude but also of the phase. Figure 2 shows diurnal variations of amplitude (upper panel) and phase (lower panel) on NSC/45.90 kHz signal against universal time (UT) recorded by AWESOME system at Belgrade station on May 4, 2012. Moments of sunrise and sunset for the receiver site and NSC transmitter are labeled by vertical lines on the Figure 2, respectively.

The diurnal change of phase is characterized by midday peak and amplitude has bigger values during night than in daytime condition, because of lower absorption. Taking into consideration the measured data shown in Figure 2, it is possible to follow nighttime, sunrise, daytime, sunset and again nighttime propagation conditions.
\end{multicols}

\centerline{\includegraphics[width=0.7\textwidth]{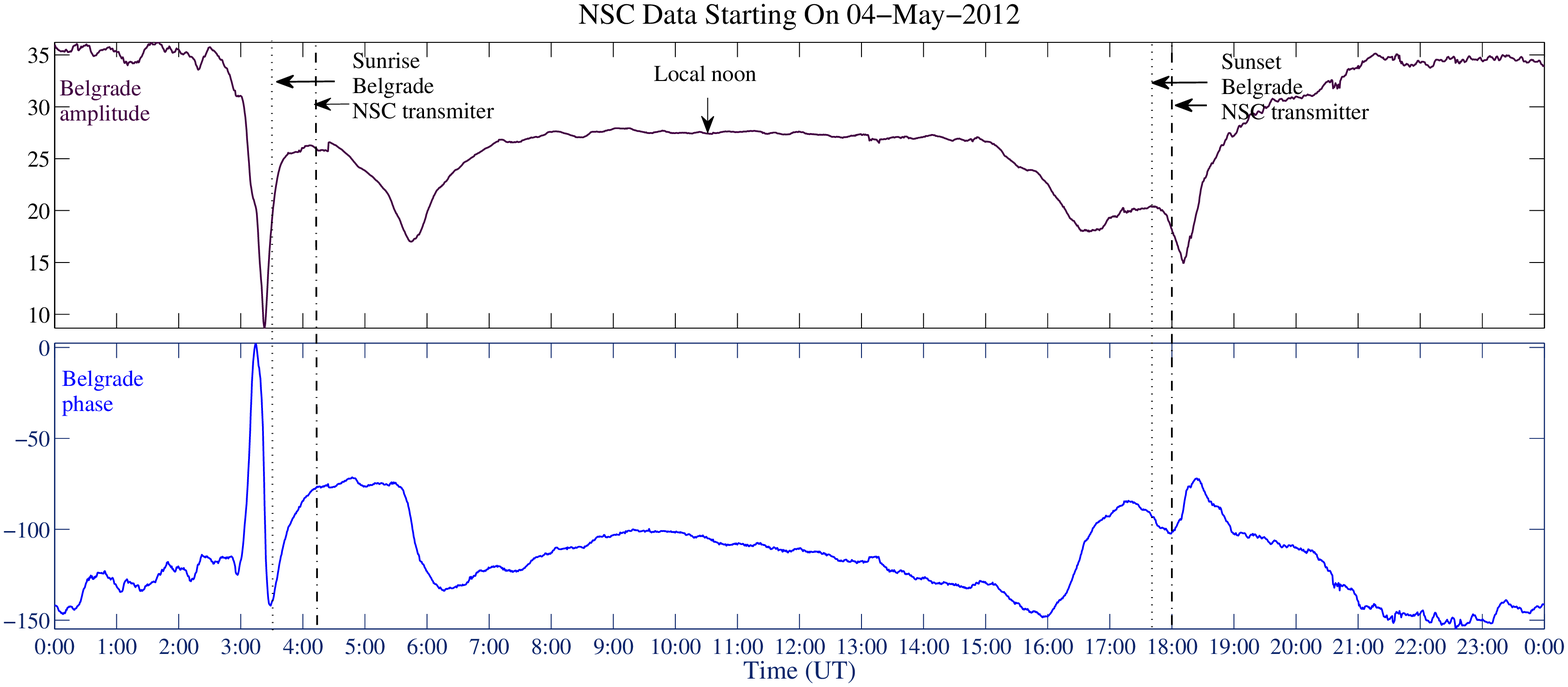}}
\figurecaption{2.}{The amplitude (upper panel) and phase (lower panel) as a function of universal time of NSC/45.90 kHz signal, as recorded at Belgrade during 4 May 2012}

\begin{multicols}{2}
\subsection{3.1. Perturbations of VLF/LF radio signals by solar X-ray flare }

Simultaneous observations of amplitude ($A$) and phase ($\Phi$) in VLF/LF signals during solar flares could be applied to calculations of electron density profile. Therefore, the perturbation of amplitude was estimated as a difference between values of disturbed amplitude induced by flare and amplitude during normal condition in the D-region: $ \Delta A=$ $A_{dis}-$ $A_{nor}$, where $dis$ means disturbed and $nor$ means normal condition. In the same way perturbation of phase was estimated as: $ \Delta \Phi=$ $\Phi_{dis}-$ $\Phi_{nor}$.

Among 200 events of amplitude and phase perturbations on ICV/20.27 kHz and NSC/45.90 signals recorded at Belgrade station there is one induced by minor B8.8 class solar flare and one induced by extremely large X1.44 class solar flare.  This large solar flare occurred on July 12, 2012 with maximum at 16:49 UT. All other examined perturbations of amplitude and phase on ICV/20.27 kHz and NSC/45.90 kHz signals were induced by small and moderate solar flares.
\end{multicols}

\centerline{\includegraphics[width=0.35\textwidth,height=0.38\columnwidth]{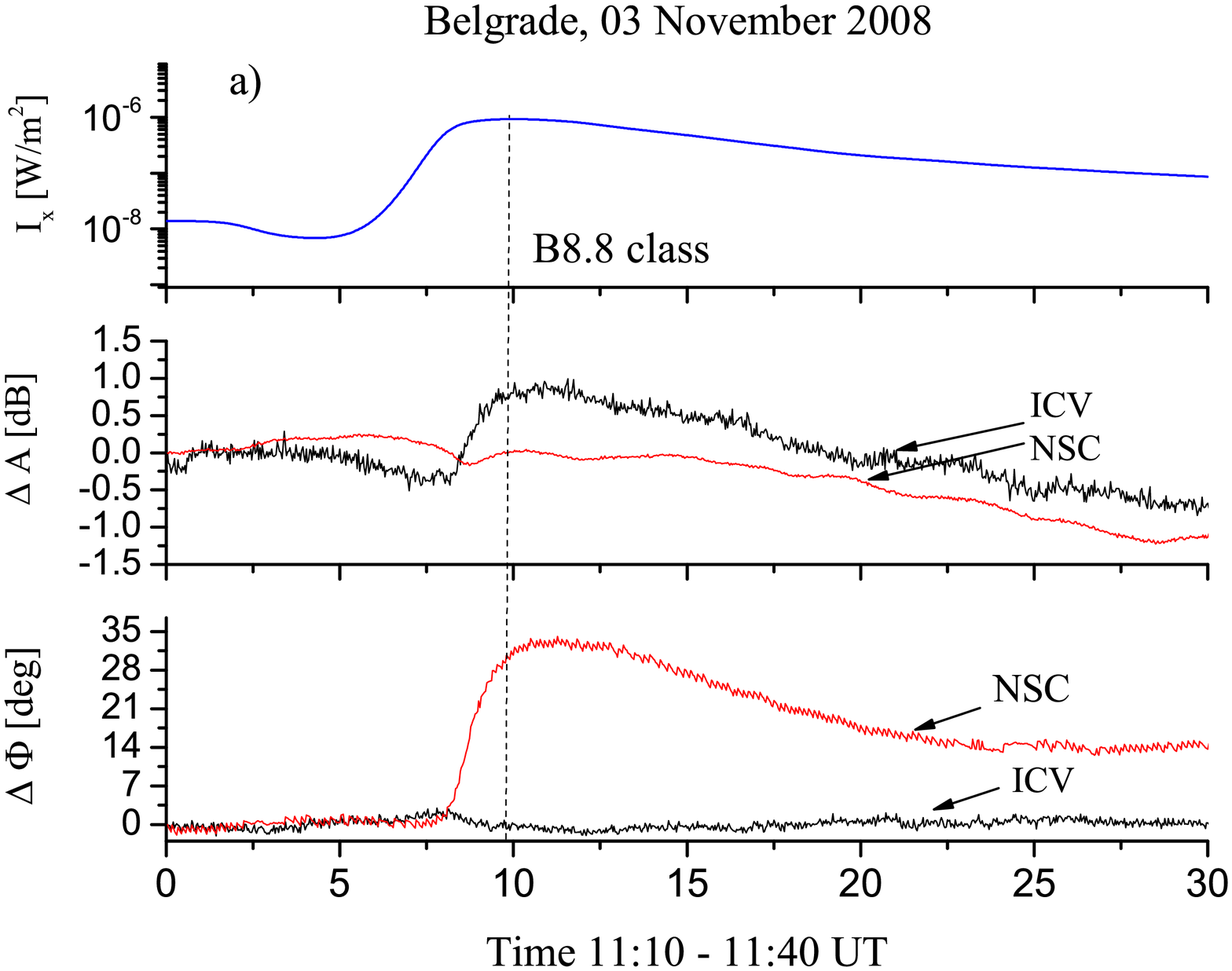}
\includegraphics[width=0.35\textwidth,height=0.38\columnwidth]{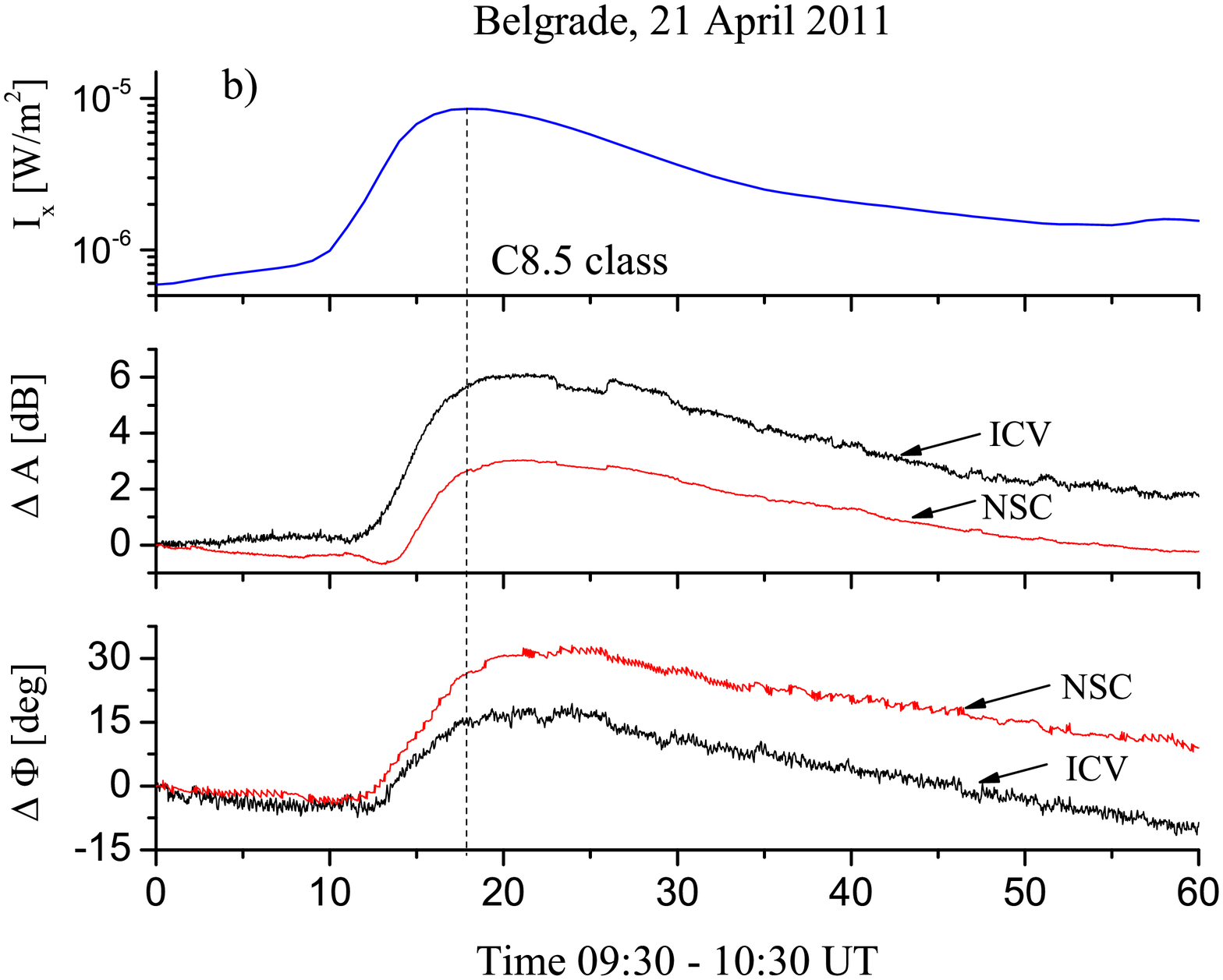}
\includegraphics[width=0.35\textwidth,height=0.38\columnwidth]{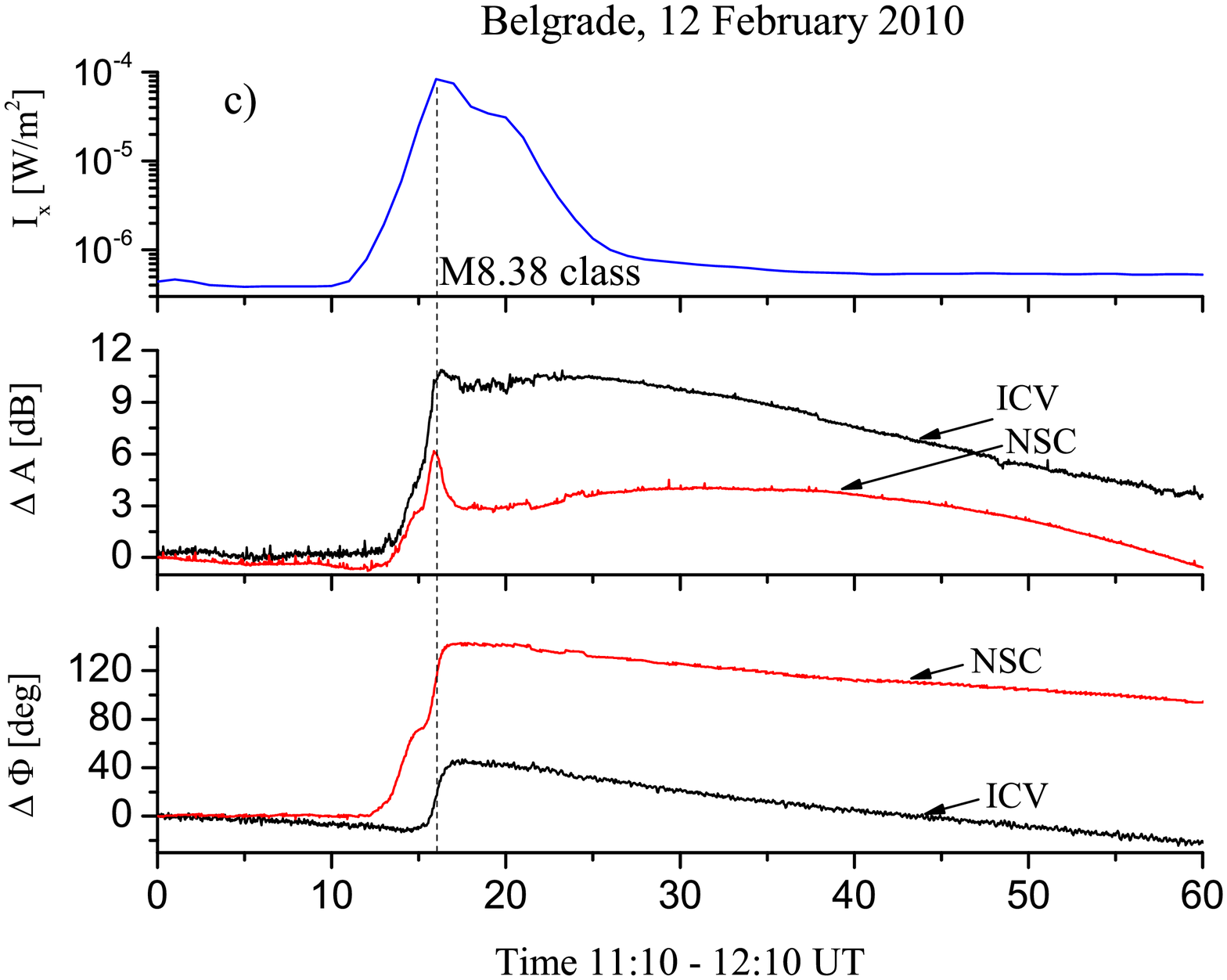}}
\figurecaption{3.}{Perturbations of amplitude and phase on NSC/45.90 kHz and ICV/20.27 kHz signals recorded at Belgrade station during minor \textbf{B8.8} class solar flare, small \textbf{C8.5} class solar flare and medium \textbf{M8.38} class solar flare as shown in the panels a), b) and c), respectively.}

\begin{multicols}{2}
On November 3, 2008 B8.8 class solar X-ray flare occurred with maximum at 10:20 UT. Although B-flares are considered minor, the blast nevertheless made itself felt on the D-region. Figure 3a shows time variation of X-ray irradiance, measured disturbances in amplitude and phase ($\Delta A$ and $\Delta \Phi$) on ICV/ 20.27 kHz and NSC/45.90 kHz signals recorded at Belgrade station on November 3, 2008. On the upper panel is shown time variation of X-ray irradiance. Disturbances in amplitude and phase on both signals caused by solar X-ray flare are shown on the middle and lower panel, respectively. Perturbation of amplitude on ICV/20.27 kHz signal had an increase of $\Delta A$ = 0.9 dB. There was no evidence of perturbation on phase. There was not recorded perturbation of amplitude on NSC/45.90 kHz signal, while at the same time phase increased by $\Delta \Phi$ $\sim 30^{0}$. Detecting a minor B8.8 flare is a good test of the method sensitivity of monitoring solar activity. Also it is the first evidence of amplitude and phase sensitivity to the influence of solar flare in a function of the transmitter frequency.

In Figure 3b there are X-ray irradiance, perturbations of amplitude and phase on ICV/20.27 kHz and NSC/45.90 kHz signals during small C8.5 class solar flare on April 21, 2011. For this a bit stronger flare additional ionization in the D-region makes disturbances of both amplitude and phase signals of ICV/20.27 kHz and NSC/45.90 kHz. As it is presented in Figure 3b the shapes of perturbed amplitude (middle panel) and phase (lower panel) on VLF and LF signals are very similar to each other. This similarity could be identified in all disturbances on ICV/20.27 kHz and NSC/45.90 kHz signals caused by small solar flares.

On February 12, 2010 M8.38 class solar flare occurred with maximum intensity of $I_{X} = 8.38\cdot10^{-5}$ W/m$^2$ at 11:26 UT and produced intensive SID. Figure 3c shows time variation of X-ray irradiance (upper panel). Measured disturbances in amplitude and phase on ICV/20.27 kHz and NSC/signal are presented on middle and lower panel of Figure 3c. The increase of amplitude on ICV/20.27 kHz was very high and at its maximum it had value higher for 10.3 dB with respect to the normal level before beginning solar flare. Simultaneously the phase of signal was disturbed with maximum increase of $\Delta \Phi$ = 44$^{0}$.
The intensity of X-ray suddenly increased and after reaching maximum slowly decreased, simultaneously the amplitude on NSC/45.90 kHz signal passed through the peak and a few minutes after the moment of maximum intensity X- ray amplitude slowly increased again to the another peak. Disturbance of phase was manifested with increase of $\Delta \Phi$ = 138$^{0}$ at 10:28 UT and after that slowly decreased to normal daytime level of phase.

Table 1 provides numerical values of perturbations amplitude and phase on ICV/20.27 kHz and NSC/40.95 kHz signals recorded at Belgrade station induced by different solar X-ray flares types.

\end{multicols}
\textit{{\bf Table 1.} Numerical values of perturbations amplitude and phase on VLF and LF signals induced by different solar flares}

\textit{}

\begin{tabular}{|p{0.8in}|p{0.5in}|p{0.7in}|p{0.5in}|p{0.7in}|p{0.5in}|p{0.7in}|} \hline
 & \multicolumn{2}{|p{1.2in}|}{Minor solar flare\newline \textbf{B8.8} class\newline $I_X=8.8\cdot10^{-7}$ [W/m$^2$]} & \multicolumn{2}{|p{1.2in}|}{Small solar flare\newline \textbf{C8.5} class\newline $I_X=8.5\cdot10^{-6}$ [W/m$^2$]} & \multicolumn{2}{|p{1.2in}|}{Moderate solar flare\newline \textbf{M8.38} class\newline $I_X = 8.38\cdot10^{-5}$ [W/m$^2$]} \\ \hline
 & $\Delta A$ [dB] & $\Delta \Phi$  [deg] & $\Delta A$ [dB] & $\Delta \Phi$  [deg] & $\Delta A$ [dB] & $\Delta \Phi$  [deg] \\ \hline
\textbf{ICV/20.27 kHz} & 0.9 & 0$^{0}$ & 6.2 & 15$^0$ & 10.3 & 44$^0$ \\ \hline
\textbf{NSC/45.90 kHz} & 0 & $30^0$ & 3.76 & 40$^0$ & 6.63 & 138$^0$ \\ \hline
\end{tabular}
\begin{multicols}{2}

In all the examined events, the amplitude of ICV/20.27 kHz signal is more sensitive than the phase due to the SIDs (see, e.g. Fig.3). Figure 4a shows the measured excess of perturbation amplitude, $\Delta A=$ $A_{dis}-$ $A_{nor}$, on ICV/20.27 kHz signal recorded at Belgrade station as a function of X-ray irradiance. The examined events were recorded in time sector for zenith angle $\chi \leq 60^{0}$. In Figure 4 dots present the measured $\Delta A$ during small and moderate solar X-ray flares. Measured excess of amplitude has value  $1 \leq \Delta A \leq 10$. Figure 4a shows that $\Delta A$ is nearly proportional to the logarithm of the maximum of X-ray irradiance, with the Adjusted R-Square and Pearson's r coefficients equal 0.83 and 0.91, respectively.

Measured excesses of amplitude $\Delta A$ on ICV/20.27 kHz signal induced by moderate solar X-ray flares are larger then $\Delta A$ on GQD/20.21 kHz and NAA/24.0 kHz signals also recorded at Belgrade station. VLF signal propagates from GQD transmitter to receiver site over great circle distance of 2000 km and VLF signal emitted from NAA transmitter propagates to receiver site over great circle distance of 6560 km. The results about propagation characteristics and perturbations amplitude and phase on those VLF signals are published by Zigman et al., (2007) and Grubor et al., (2008).

The flare induced phase perturbations always increase on ICV/20.27 kHz signal.
\end{multicols}

\centerline{\includegraphics[width=0.5\textwidth]{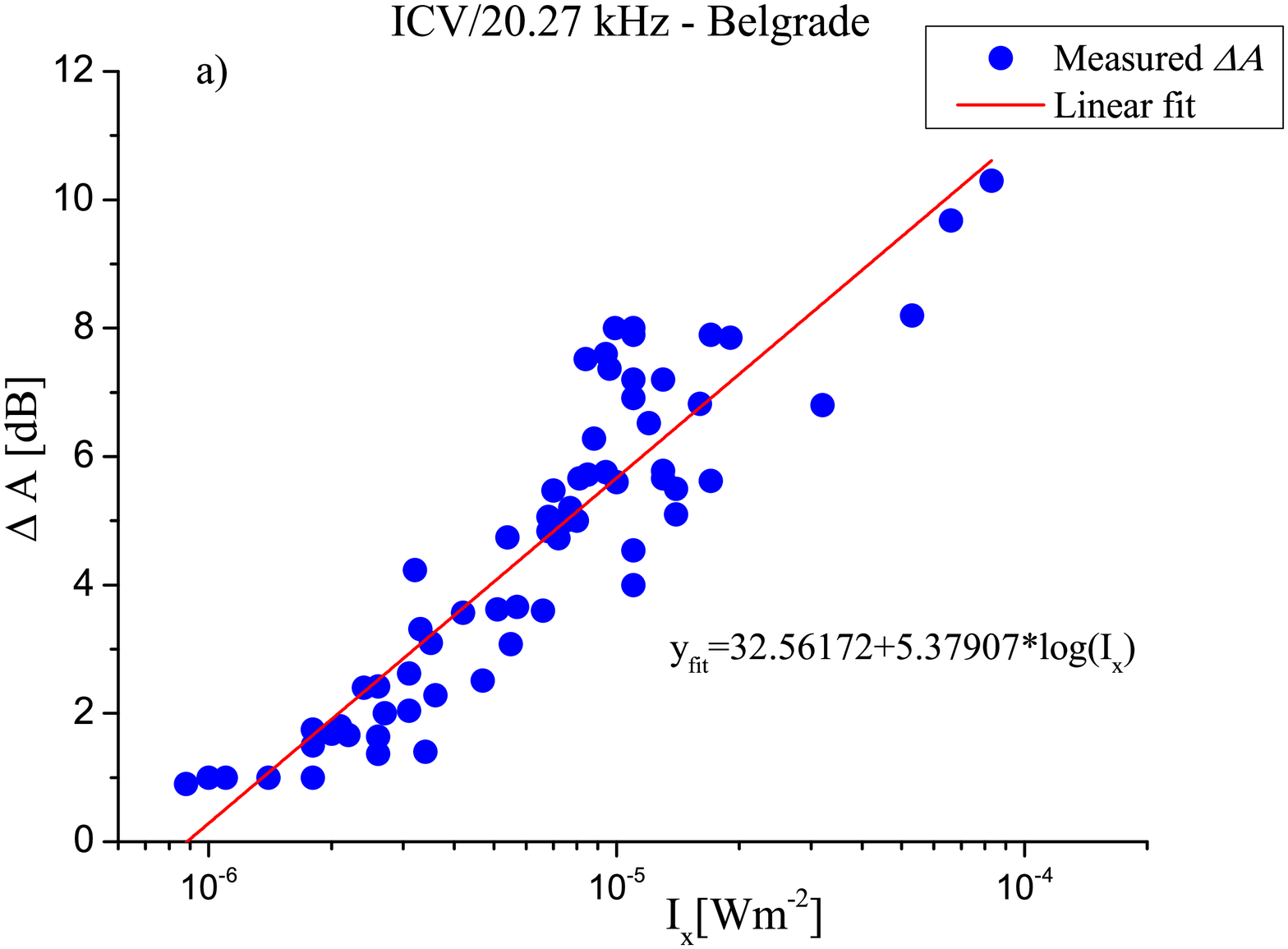}
\includegraphics[width=0.5\textwidth]{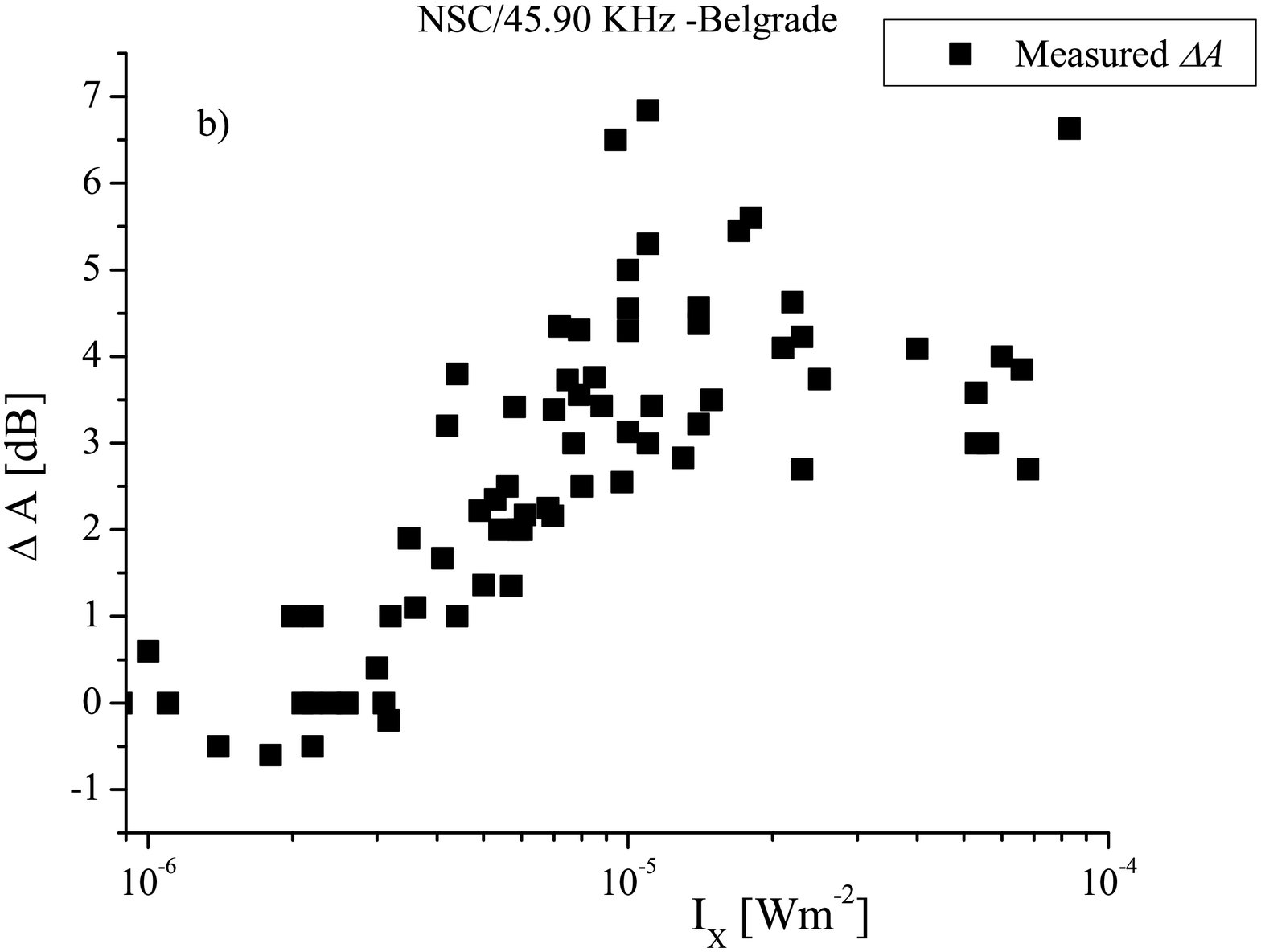}}
\figurecaption{4.}{ Measured excesses of perturbation amplitude on ICV/20.27 kHz signal as a function of logarithm solar X- ray irradiance a), and  Measured excesses of perturbation amplitude on NSC/45.90 kHz signal as a function of logarithm solar X- ray irradiance b).}

\begin{multicols}{2}

As we have pointed out in this paper the best defined amplitude and phase of signal received by AWESOME system at Belgrade station originates at the NSC transmitter from Sicily, Italy at 45.90 kHz. This LF signal propagates over great circle distance of 952 km. Perturbations of amplitude are not well defined during solar flares occurrences with X-ray irradiances, less than irradiance corresponding to C3 class solar flare. Measured excess of amplitude has value $-0.6 \leq \Delta A \leq 1$ dB. In Figure 4b squares present the measured $\Delta A$ during small and moderate solar X-ray flares. Solar flares from C3 class to approximately M2 class caused perturbations of amplitude and phase on signal. Variation of perturbed amplitude against time is very similar to time variation of X-ray irradiance. The result of our study is that solar flares larger than M2 class induced perturbation of amplitude in way that amplitude oscillates around the level before beginning of the flare.  The amplitude is perturbed in the way that it passes through peak and then trough next peak, etc. In all examples that we analyzed the amplitude reaches the first peak in time close to maximum intensity of solar X-ray flare. After that the amplitude decreases during next few minutes reaching the minimum. This is the consequence caused by complicated nature of physical processes in the perturbed D-region and LF signal propagation.

Recorded phase of NSC signal is very sensitive to SIDs induced by solar X-ray flare. Perturbation of signal phase has time variation very similar to time variation of solar flare irradiance. An increase of phase induced by small solar X-ray flare reached maximum at moment that is very close to time of maximum intensity of solar flare. When moderate solar X-ray flare induces disturbances in the D-region, time of phase maximum is few minutes after moment of maximum intensity of solar X-ray flare.

\subsection{3.2. Reproducibility of the measured amplitude perturbations.}

Solar flare affects the recorded VLF/LF signal amplitude and phase. The stability and reproducibility of the received amplitude and phase on NSC/45.90 kHz signal were found in many examined events recorded during the period of six years. To illustrate that, Figure 5a provides time variation of perturbed amplitudes affected by two C class solar flares that occurred about a year apart. One solar flare occurred on May 5, 2010 with peak irradiance $I_{X} = 8.8\cdot10^{-6}$ W/m$^{2}$ and the other on April 21, 2011 with peak irradiance $I_{X} = 8.5\cdot10^{-6}$ W/m$^{2}$. Time variations of perturbed amplitude $A$ and amplitude excess $\Delta A$ for these two events have almost identical shapes and as can be seen even the values.

Figure 5b shows reproducibility of amplitude perturbations affected by M class solar flares. On July 2, 2012 there was a flare with peak irradiance $I_{X} = 5.6\cdot10^{-5}$ W/m$^{2}$ and on March 8, 2011 the flare with peak irradiance of $I_{X} = 5.3\cdot10^{-5}$ W/m$^{2}$ was recorded. Both of them affected the amplitudes. Time variations of perturbed amplitude for these two events have very similar shapes.

\end{multicols}

\centerline{\includegraphics[width=0.5\textwidth]{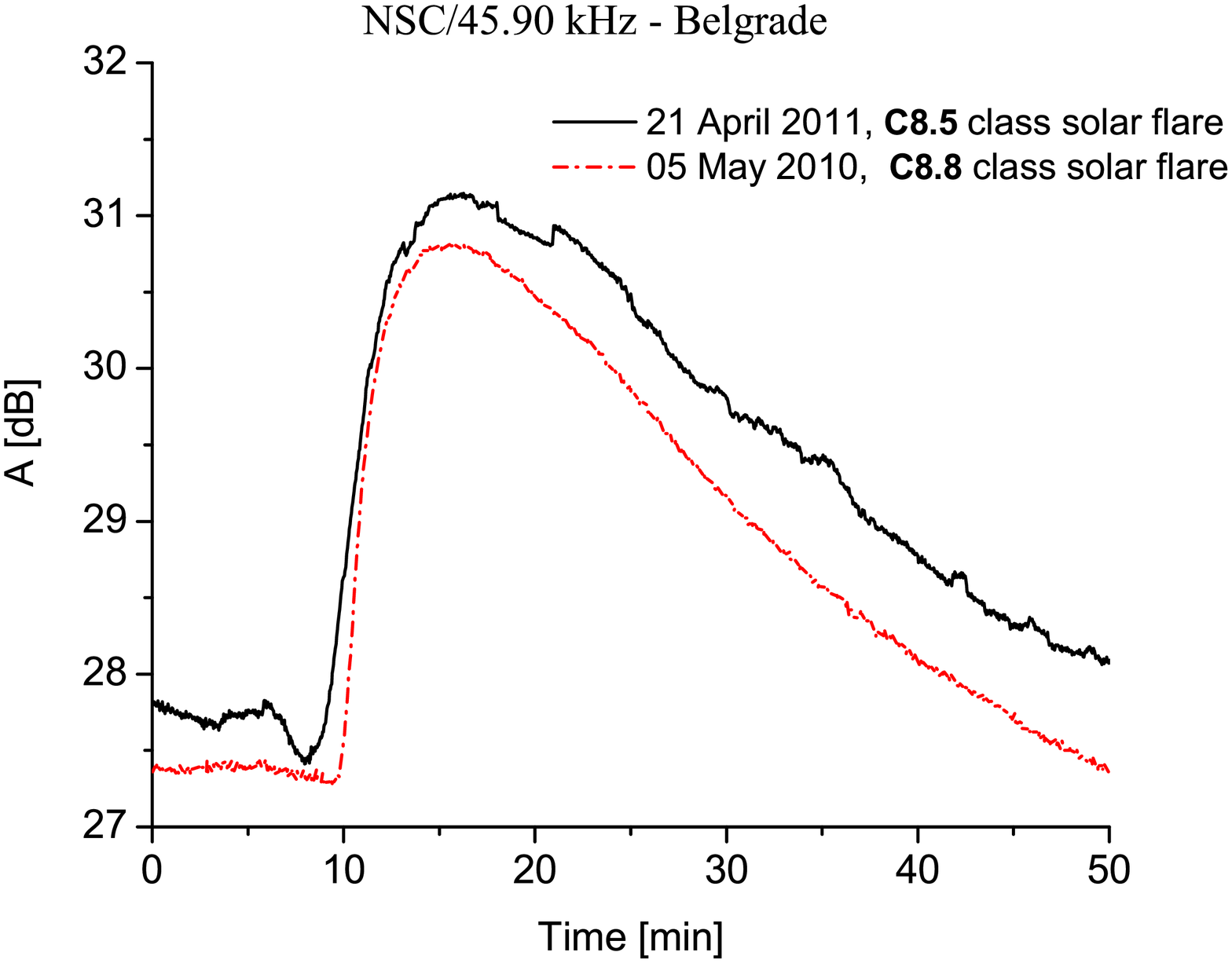}\includegraphics[width=0.5\textwidth]{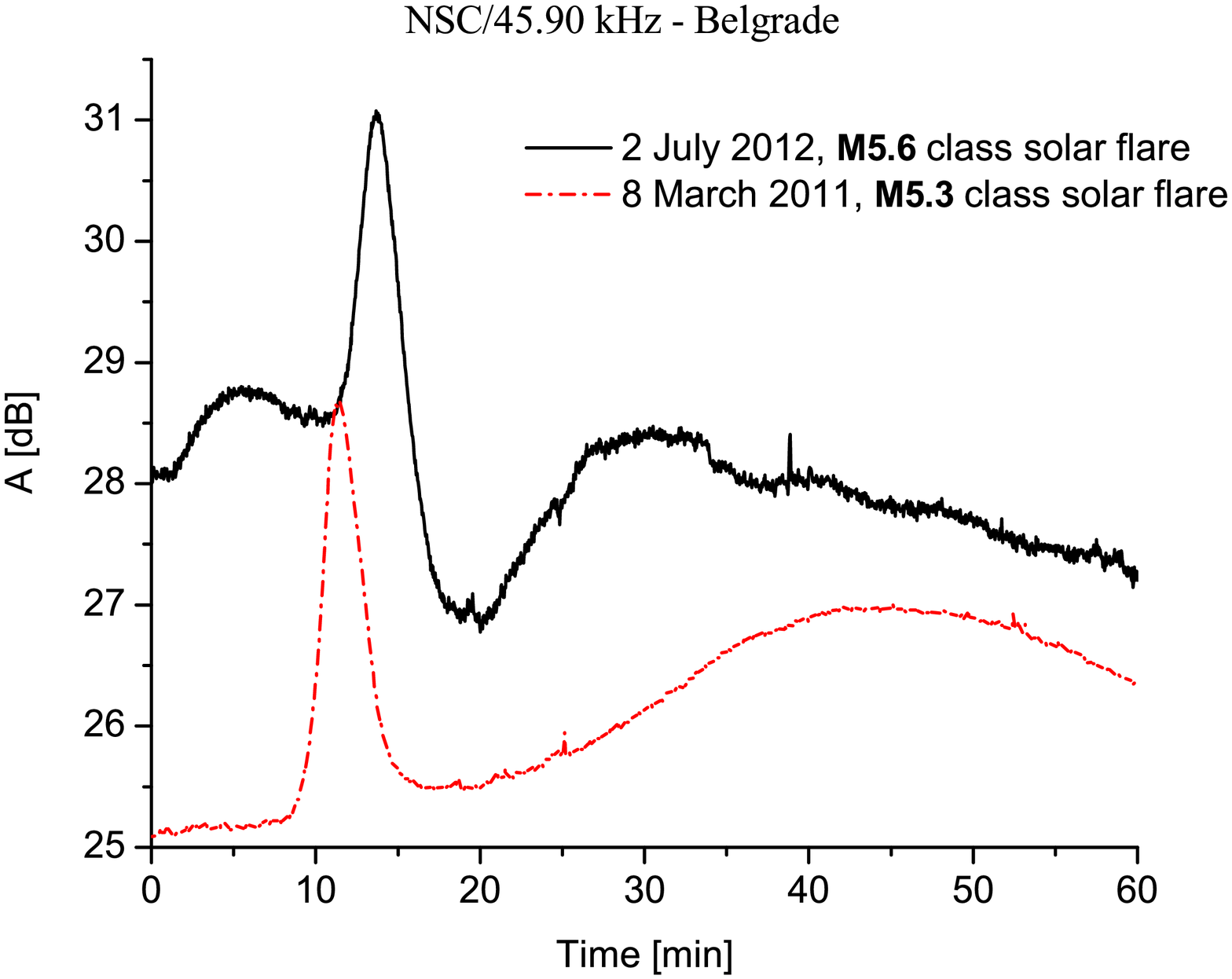}}
\figurecaption{5.}{Reproducibility of the measured amplitude perturbations. Parallel presentation of two events of amplitude perturbation on signal NSC/45.90 kHz versus time: a) induced by two C class flares which occurred in 05 May 2010 and 21 April 2011; b) induced by two M class flares, occurred in 2 July 2012 and 8 March 2011.}

\begin{multicols}{2}

Similar level of X-ray irradiances makes changes in the lower ionosphere, which also similarly influences the characteristics of subionospheric VLF/LF propagation as an increase in amplitudes. This reproducibility and their sensitivity make propagation of VLF/LF signals a useful and reliable tool in evaluating electron density in the D-region.

\section{4. Method of simulation: application to solar flare induced perturbations of VLF/LF radio signals}

Radio signals in VLF and LF ranges from transmitters propagate through waveguide bounded by the Earth's surface and D-region. This propagation is stable both in amplitude and phase and has relatively low attenuation. Characteristics of the D-region have strong influence on propagation of VLF/LF signals. Theoretical base for this propagation under normal, undisturbed ionospheric conditions is developed by Wait and Spies (1964). The influence of the D-region is taken in account by using the so called Wait's parameters: sharpness beta(km$^{-1}$) and reflection height $h^{'}$(km). The Naval Ocean System Center, San Diego, USA has developed computer program LWPC- Long Wave Propagation Capability, for simulation of VLF/LF propagation along any particular great circle path under different diurnal, seasonal and solar cycle variations in the ionosphere (Ferguson, 1998). The LWPC program can take arbitrary electron density versus altitude profiles supplied by the user to describe the D-region and thus the ceiling of the waveguide.

The Wait's parameters were used successfully to calculate electron density height profiles in the D-region under regular change of day/night and season or solar cycle variation (Thomson, 1993; Thomson and Clilverd, 2001; McRae and Thomson, 2000). A very marked change in the electron density is caused by solar X-ray flares, so the Wait's parameters were used in calculations of electron density height profiles in suddenly disturbed D-region, (Thomson and Clilverd, 2001; Thomson, 2005; Zigman, et al., 2007; Nina et al. 2011, 2012a, 2012b; {\v S}uli{\'c} et al., 2010; Grubor, et al., 2008).

A convenient quantity to describe the characteristics of the D-region is based on a \textit{height dependent conductivity parameter}, $\omega_{r}(h)$. Wait and Spies defined this quantity as: $\omega_{r}(h)=\omega_{p}^{2}(h)/\nu(h)=2.5\cdot10^{5}\cdot e^{\beta\cdot(h-h')}$ where $\omega_{p}$ is the electron plasma frequency (angular) and $\nu(h)$ is the effective electron-neutral collision frequency, both being functions of the height, $h$, in km. This definition assumes that $\omega_{r}(h)$ varies exponentially with height at a rate determined by the constant $\beta$. $h'$ is the height (km) at which $\omega_{r}(h)=2.5\cdot10^{5}$ rad/s and has been found to be convenient measure of the "reflection height" of the ionosphere (Wait and Spies, 1964).
The electron density profile increases exponentially with height and can be associated with above defined equations.
The equation for the electron density in the D-region:
\begin{equation}
\label{eq:Ne}
N_{e}(h,t)=1.43\cdot10^{13} e^{0.15h'(t)} e^{(\beta(t)-0.15)\cdot(h-h'(t))} \quad \text{m}^{-3},
\end{equation}
was developed by Thomson (1993) and we also use it in our work to calculate the vertical density profile in the range 50 - 90 km.

Typically the electron density profile is changing from day to day even during magnetically quiet periods. Our intention was to calculate electron density profile versus height, and versus time. To estimate sharpness $\beta(\text{in km}^{-1})$ and reflection height $h'$ for normal condition in daytime D-region we used below formulas developed by Ferguson:

\begin{equation}
\label{eq:beta}
\begin{split}
\beta_{0}=0.5349-0.1658\cos\chi -0.08584\cdot\\
\cdot\cos\varphi +0.1296X_{5} \quad \text{km}^{-1}
\end{split}
\end{equation}
\begin{equation}
\label{eq:h'}
\begin{split}
h'_{0}=74.37-8.087\cos\chi+5.779\cos\theta\\
-1.213\cos\varphi-0.0044X_{4}-6.035X_{5} \quad \text{km}.
\end{split}
\end{equation}
Where: $\chi$ is solar zenith angle, $\theta$ is geographical latitude, $\varphi=2\pi (m-0.5)/12$,  $m$ is ordinal number of month in the year, $X_{4}$ is Zurich sunspot number and $X_{5}$ = 0 or 1 for magnetically quiet or disturbed conditions, respectively (Ferguson, 1980)

The first step of used \textit{trial and error method} is the simulation of VLF/LF signals propagating from transmitter to the receiver site under normal ionospheric condition in D-region and to estimate sharpness $\beta_{0}$ and reflection height $h'_{0}$ by Eqs's (1) and (2). Using LWPC program with those Wait's parameters we simulated propagation of VLF/LF signals and obtained simulated values for amplitude and phase.

The next step is to simulate propagation of VLF/LF signals through waveguide while D-region was disturbed for the characteristics moments during the flare development. We choose the next pair of $\beta$ and $h'$ and used it as an input parameter in LWPC program to obtain simulated values of amplitude and phase at the receiver site.

The process was repeated until differences between simulated amplitudes and phases for disturbed and normal condition in the D-region matched with measured differences $\Delta A$ and $\Delta \Phi$, respectively. After good matching: $\Delta A_{sim}$  $\approx \Delta A$ and $\Delta \Phi_{sim}$  $\approx \Delta \Phi$, that pair of $\beta$ and $h'$ was used for calculation of electron density profile versus height for characteristic moments during and after solar flare occurrence. In the used expressions index "sim" means simulated.

\subsection{4.1. Example of an electron density profile in the perturbed D-region}

On February 2, 2014 occurred seventeen solar flares. In this paper we presented results of perturbed electron density in D-region induced by M4.4 class solar flare with peak irradiance $I_{X}=4.46\cdot 10^{-5}$ Wm$^{-2}$ at 09:31 UT. This flare event caused disturbances of the amplitude and phase on NSC/45.90 signal as is evident from the plots given in Figure 6. Figure 6a gives X-ray irradiance and observed perturbation amplitude on LF signal versus universal time. As X-ray irradiance increased it caused increase of amplitude till 09:29 UT, when the amplitude decrease began. In time interval 09:29 to 09:32 UT amplitude passed through a minimum. This time interval corresponded to time of X-ray peak irradiance. Sudden disturbances in D-region changed propagation characteristics of LF radio signal, which resulted in process of signal attenuation.

Figure 6b shows time variation of X-ray irradiance and measured phase on signal for time interval 09:15 - 10:15 UT. Time variation of perturbed phase is in close correlation with time variation of the X-ray irradiance. The maximum of perturbed phase occurred at 09:32 UT, lasting one minute after appearance of X-ray peak irradiance.

\end{multicols}

\centerline{\includegraphics[width=0.5\textwidth]{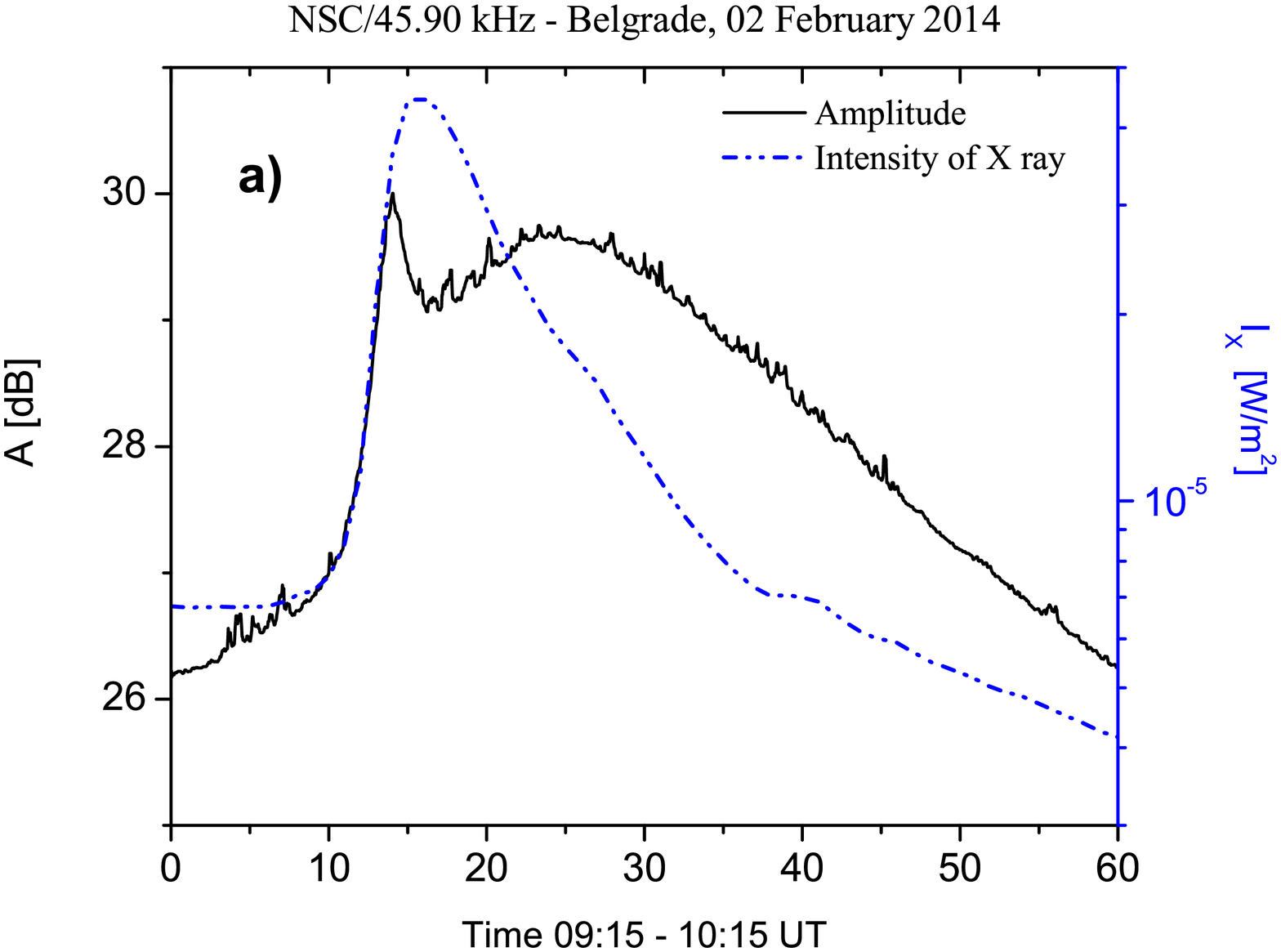}
\includegraphics[width=0.5\textwidth]{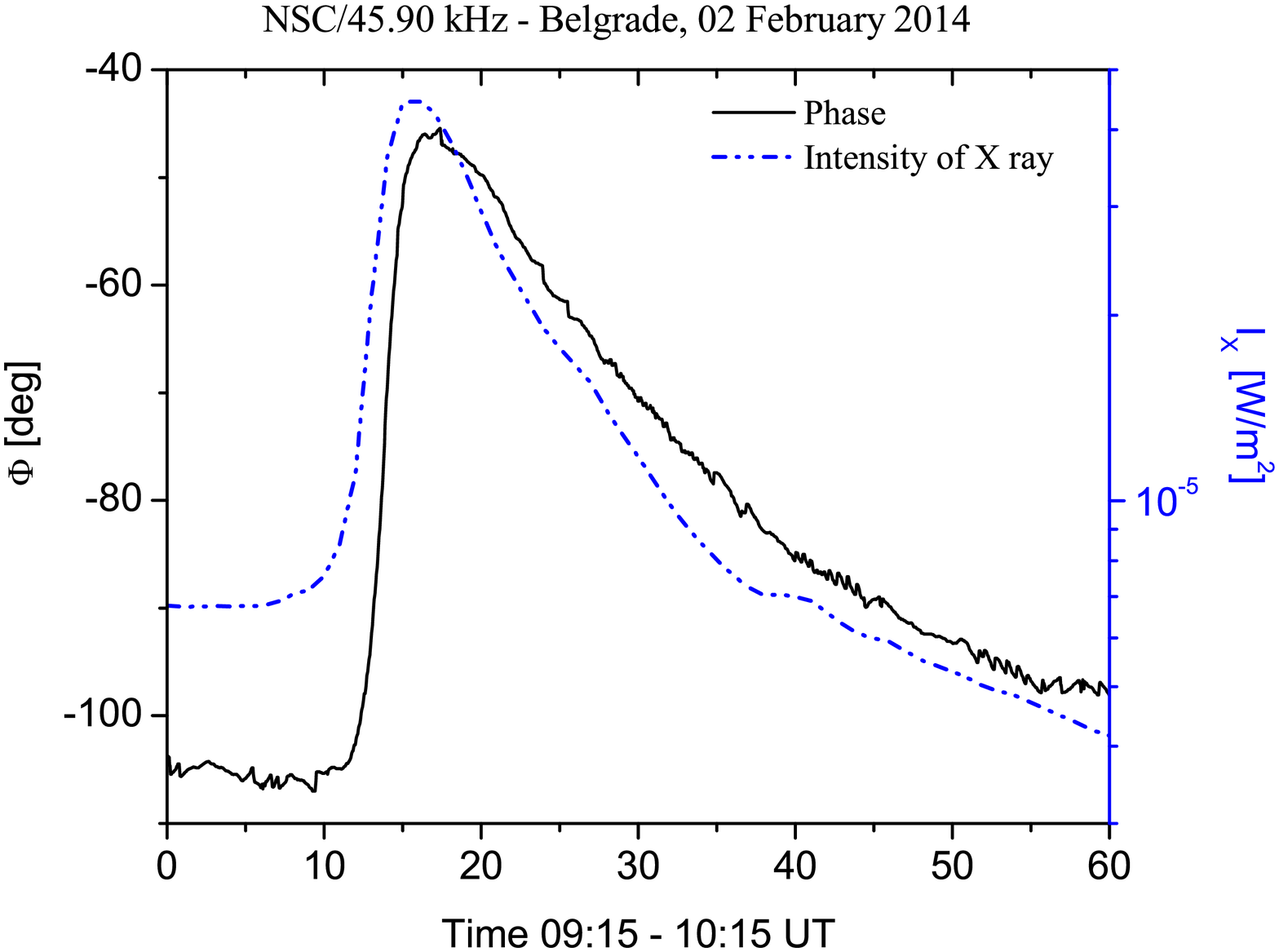}}
\figurecaption{6.}{Time variation of X-ray irradiance measured by GOES-15 satellite (dash dot dot), and perturbed amplitude and phase
of signal emitted from NSC transmitter and recorded on AWESOME receiver in Belgrade (Serbia) during observed M4.46 class solar
flare (from 09:15 to 10:15 UT) on February 2, 2014.}

\begin{multicols}{2}
One of the purpose of this work was to calculate electron profile in a function of time, during and after occurrence of solar flare.  In accordance to this, it was necessary to obtain $\beta(t)$ and $h'(t)$, during the whole examined time interval, as input parameters for LWPC program.

\end{multicols}

\textit{{\bf Table 2.} Values of measured amplitude and phase perturbations and evaluated quantities $\beta$ and $h'$ and electron densities $N_{e}$ (at h=70km) for different moments during and after occurrence of M4.46 class solar flare.}
\vskip.25cm

\begin{tabular}{|c|c|c|c|c|c|c|} \hline
Time [UT] & $I_{X}$  [$10^{-5}$ Wm$^{-2}$]  & \multicolumn{5}{|c|}{ NSC/45.90 kHz} \\ \hline
 &  & $\Delta A$ [dB] & $\Delta \Phi$ [deg]  & $\beta$ [km$^{-1}$] & $h'$ [km]  & $N_{e}$ [$10^{9}$m$^{-3}$]  \\ \hline
09:15 & 0.677 & -     & -   & 0.350 & 70.00 & 0.39 \\ \hline
09:24 & 1.900 & 0.60  & -3  & 0.380 & 69.20 & 0.53 \\ \hline
09:27 & 1.550 & 1.25  & 28  & 0.397 & 68.20 & 0.80 \\ \hline
09:29 & 3.600 & 3.65  & 32  & 0.403 & 65.70 & 2.20 \\ \hline
09:32 & 4.260 & 3.05  & 58  & 0.357 & 64.55 & 2.76 \\ \hline
09:35 & 2.950 & 3.24  & 54  & 0.358 & 64.90 & 2.40 \\ \hline
09:40 & 1.770 & 3.42  & 42  & 0.365 & 65.70 & 1.89 \\ \hline
09:45 &	1.800 & 3.18  & 33	& 0.362	& 66.55	& 1.37 \\ \hline
09:50 &	0.803 & 2.57  & 26	& 0.359	& 67.30	& 1.04 \\ \hline
09:55 &	0.699 & 2.08  & 18	& 0.352	& 68.15 & 0.76 \\ \hline
10:00 &	0.599 & 1.51  & 14	& 0.352	& 68.65	& 0.63 \\ \hline
10:05 &	0.529 & 0.96  & 10	& 0.350	& 69.04	& 0.55 \\ \hline
10:10 &	0.283 & 0.48  & 6	& 0.350	& 69.50	& 0.47 \\ \hline
\end{tabular}

\begin{multicols}{2}
In Table 2 there are measured values: $\Delta A$, $\Delta \Phi$, together with quantities obtained by LWPC program: sharpness, $\beta$, reflection height $h'$ and electron density $N_{e}$ at referent height $h = 70$ km for different moments during M4.46 class solar flare occurrence and recovery period.
The X-ray irradiance $I_{X}=4.26\cdot 10^{-5}$ Wm$^{-2}$ at 09:32 UT induced amplitude and phase perturbations of $\Delta A$ = 3.05 dB and $\Delta \Phi = 58^{0}$, respectively. We obtained for sharpness $\beta$ = 0.357 km$^{-1}$ and for reflection height $h'$ = 64.55 km which means that the reflection height moved lower for 5.45 km.  The electron density increased at $h$ = 70 km from 3.94$\cdot 10^{8}$ m$^{-3}$ to 2.76$\cdot 10^{9}$ m$^{-3}$.

\centerline{\includegraphics[width=\columnwidth,
height=0.85\columnwidth]{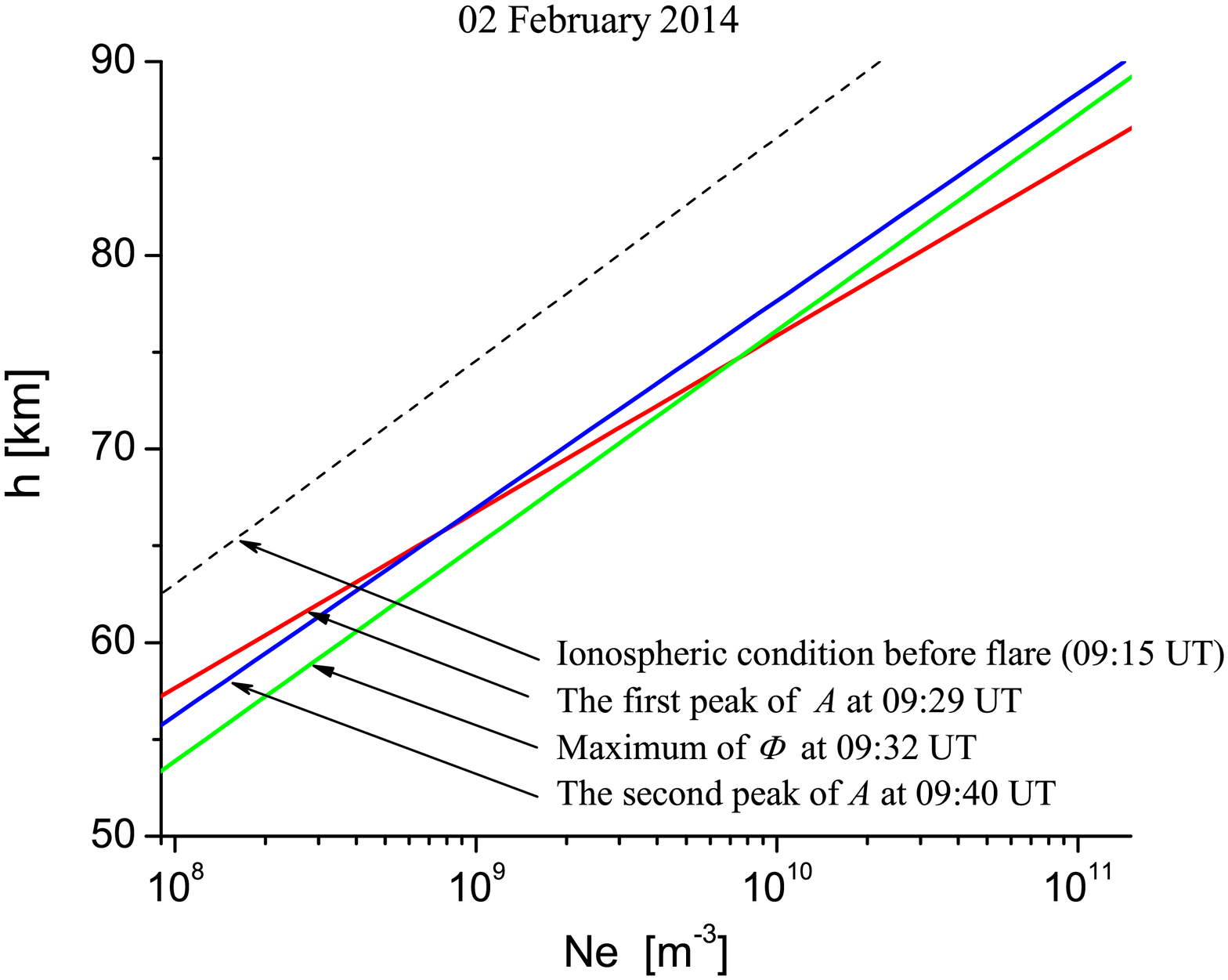}}
\figurecaption{7.}{The flare time electron densities from the values of $\beta$ and $h'$ given in Table 2 (the height
profile of electron density).}
Figure 7 shows the vertical electron density profile before and during the M4.46 class solar flare on February 2, 2014. So, there is the vertical electron density profile before beginning of the flare (09:15UT), for the characteristics moments when amplitude passed through the first and the second peak and phase passed the maximum. The corresponding electron density profiles moved to higher electron densities with different slopes as compared to density profile at ionospheric condition before beginning of the flare.
\centerline{\includegraphics[width=\columnwidth,
height=0.82\columnwidth]{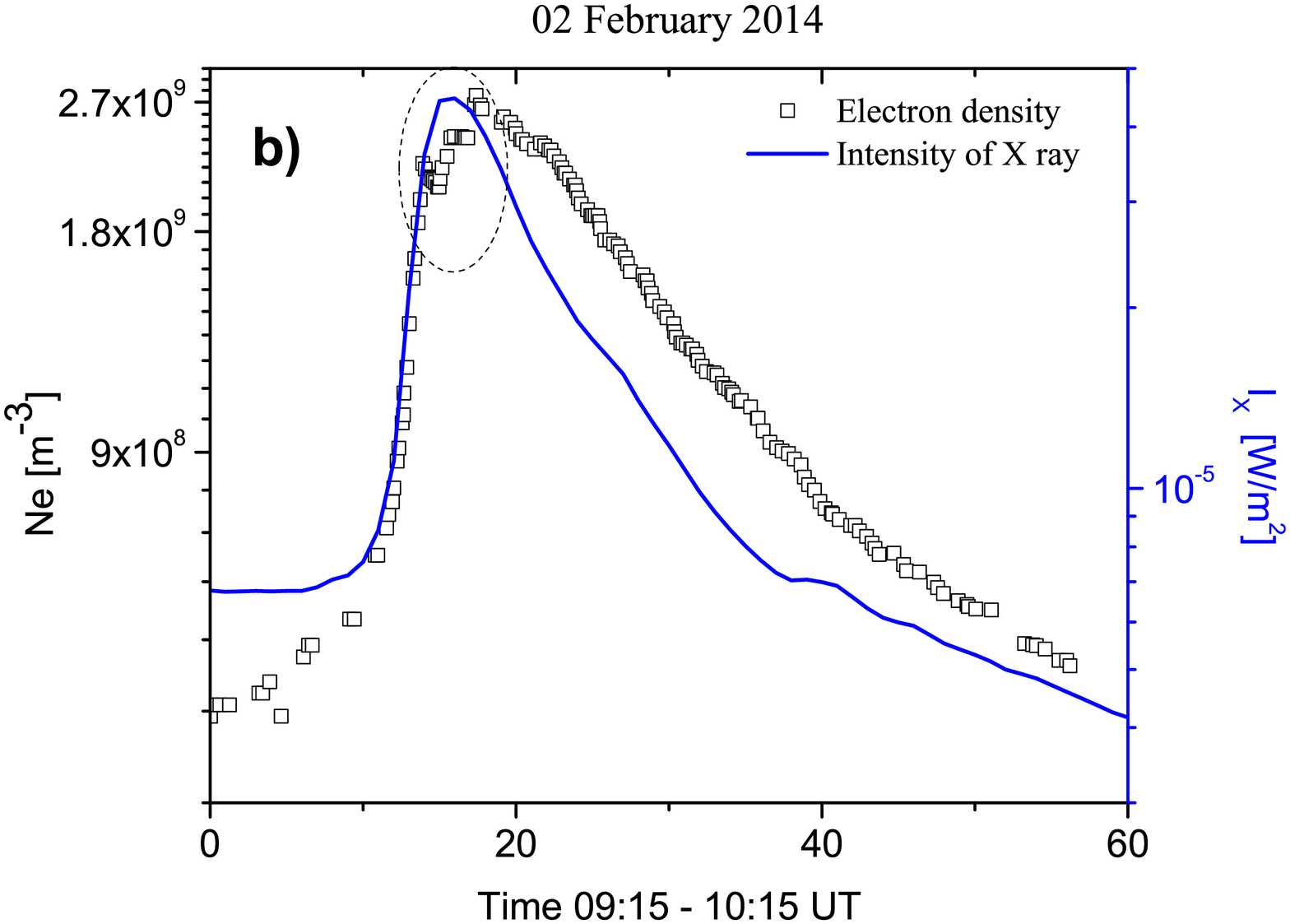}}
\figurecaption{8.}{Variation of X-ray irradiance, as measured by GOES-15 satellite, and the corresponding electron density evaluated at height 70 km versus universal time. The electron density profile is obtained on base of measured perturbed amplitude and phase on NSC/45.90 kHz signal at Belgrade station.}

Figure 8 shows simultaneous X-ray irradiance and calculated electron density (at reference height $h = 70$ km) as a function of time (for interval 09:15UT - 10:15UT). Oval marks the values of electron density based on values of attenuated amplitude after the first peak presented in Fig.6a. It can be noticed that the time distribution of the electron concentration follows the time variation pattern of the registered solar flux on GOES-15 satellite.

\section{5. DISCUSSION AND CONCLUSION}

The purpose of this work was to examine perturbations of amplitude and phase on VLF/LF radio signals propagating over great circle path with distances smaller than 1000 km, so called short path. The receiver at Belgrade station continuously monitor the amplitude and phase of coherent and subionospherically propagating radio signals operated in Sicily, NSC at 45.90, kHz and in Isola di Tavolara ICV at 20.27 kHz, with great circle distances of 953 km and 976 km, respectively. Geographically and in according to conductivity properties these two short paths are vary similar to each other. The main difference is in the transmitter frequency, and for that reason was done detailed study around 200 events of SID and their influence to amplitude and phase on ICV/20.27 kHz and NSC/45.90 kHz signals (see Fig.4). Measured data were obtained during period of six years covering minimum and maximum of solar activity.

The first our result is that the amplitude of ICV/20.37 kHz signal is more sensitive then phase on the disturbances caused by X-ray solar flare. In all events increasing X-ray irradiance induced increase of amplitude and phase on signal. It was shown that difference between perturbed and normal amplitude changed from $\Delta A$ = 1dB up to $\Delta A$ = 10 dB. It is presented on Figure 4a, showing monotonous increase of $\Delta A$ with logarithm of the max. solar X-ray irradiance ($\Delta A$ is nearly proportional to the logarithm of $I_{x}$).

Analyzing propagation characteristics of NSC/45.90 kHz signal and perturbed amplitude and phase caused by solar X-ray flare we are able to make following conclusions.

Perturbations of phase on NSC/45.90 kHz signal are very remarkable. Minor solar flare, as B8.8 class, induced increase of phase, $\Delta \Phi$ $\sim$ $30^{0}$. Shape of perturbed phase is very similar to time variation of X-ray irradiance.

During occurrence of solar flares, classified as minor and small flare up to C3 class, amplitude on signal NSC/45.90 kHz does not have significant perturbations. Solar flare in range from C3 to M3 classes induced increase of amplitude on signal. Moderate solar flares classified larger then M3 class induced oscillation of amplitude around base level before beginning of flare (Figure 5b).

The stability and reproducibility of the received amplitude and phase on NSC/45.90 kHz signal were found in many examined events recorded during six years period. This is illustrated on Fig 5a and 5b. This reproducibility gives the possibility that only in looking measured data we are able to define class of solar X-ray flare which induced that perturbation.

As the presence of electrons in the ionospheric D-region strongly affects the VLF/LF radio wave propagation we present a method for determination
of the electron density. Using measured $\Delta A$ and $\Delta \Phi$ as very important quantities and LWPC program we calculated electron density in the D-region. Also it is possible to calculate electron density profile as a function of time, to follow increase during time interval of peak irradiance, and to study decrease of electron density during recovery period.

It can be noticed that the time distribution of the electron density follows the time variation pattern of the registered solar flux on GOES-15 satellite. In the case we analyzed, the electron concentration (at reference height) has values within the order of magnitude $10^{8}$ - $10^{9}$ m$^{-3}$ during this flare (see Tab.2). Finally, we can see (Fig.7) that the considered solar flare, as expected, causes larger increases in electron concentration at higher altitudes.


\acknowledgements{
The authors are very grateful to colleagues from AOB especially Zorica Cvetkovic for a very wide and fruitful discussion. Also, the authors are thankful to the Ministry of Education, Science and Technological Development of the Republic of Serbia for support of this work within projects 176002 and III4402
}



\references

Ferguson, J.A.: 1980, in "Ionospheric profiles for predicting nighttime VLF.LF propagation, Naval  Ocean Systems Center Tech. Rept NOSC/TR 530", NTIS Accession No ADA085399, National Technical Informations Service Springfield, VA 22161, U.S.A.

Ferguson, A. J.:1998, in "Computer Programs for Assessment of Long- Wavelength Radio Communications, Version 2.0, Technical document 3030", Space and Naval Warfare Systems Center, San Diego CA 92152-5001.

Grubor, D.P., {\v S}uli{\'c}, D. and  {\v Z}igman, V.: 2008, \journal{Ann. Geophys.}, \vol{26}, 1731-1740.

McRae, M.W. and Thomson, N. R.: 2000, \journal{J. Atmos. Sol.-Terr. Phys.}, \vol{62}, 609-618.

Mitra, A.P.: 1974 in "Ionospheric Effects of Solar Flares." eds. D. Reidel, Dordrecht, Holland.

Nina, A., {\v C}ade{\v z}, V., Sre{\'c}kovi{\'c}, V.~A., and {\v S}uli{\'c}, D.: 2011, \journal{Balt. Astron.}, \vol{20}, 609

Nina, A., {\v C}ade{\v z}, V., Sre{\'c}kovi{\'c}, V., and {\v S}uli{\'c}, D.: 2012, \journal{Nucl. Instrum. Meth. B}, \vol{279}, 110

Nina, A., {\v C}ade{\v z}, V., {\v S}uli{\'c}, D., Sre{\'c}kovi{\'c}, V. and Zigman, V.: 2012,\journal{ Nucl. Instrum. Meth. B}, \vol{279}, 106

Schunk, R. W. and Nagy, A.F.: 2000, in "Ionospheres: physics, plasma physics and chemistry", Cambridge University Press, 570 pp.

{\v S}uli{\'c}, D., Nina A., Sre{\'c}kovi{\'c}, V.: 2010, arXiv:1405.3783, \journal{Pub. Astro. Obs. Belgr.} \vol{89}, 391.

Thomson , N.R.: 1993, \journal{J. Atmos. Sol.-Terr. Phys.}, \vol{55(2)}, 173 - 184.

Thomson, N.R., Rodger, C. J. and Clilverd, M.A.: 2000,  \journal{J. Geophys. Res.} \vol{110}, A06306.

Thomson, N. R. and Clilverd, A. M.: 2001, \journal{J. Atmos. Sol.-Terr. Phys.}, \vol{63}, 1729-1737.

Wait, J. R., and K. P. Spies: 1964, in "Characteristics of the Earth-ionosphere waveguide for VLF radio waves, Technical Note 300", National Bureau of Standards, Boulder, CO.

{\v Z}igman, V., Grubor, D., and {\v S}uli{\'c} D.: 2007, \journal{J. Atmos. Sol.-Terr. Phys.}, \vol{69(7)}, 775-792.

\endreferences
\end{multicols}

\vfill\eject

{\ }



\naslov{UPOREDNA ANLIZA POREME{\CC}AJA AMPLITUDE I FAZE RADIO SIGNALA VRLO NISKIH I NISKIH FREKVENCIJA U TOKU  POJAVE SUNQEVOG FLERA}


\authors{ D. M \v Suli\' c$^1$, V. A. Sre\' ckovi\' c$^2$ }

\vskip 3mm


\address{$^1$University Union - Nikola Tesla Belgrade, Serbia}
\Email{desankasulic}{gmail.com}
\address{$^2$Institute of Physics, University of Belgrade, P.O. Box 57, Belgrade, Serbia}
\Email{vlada}{ipb.ac.rs}

\vskip.7cm


\centerline{UDK \udc}



\vskip.7cm

\begin{multicols}{2}
{


{\rrm Poreme{\cc}aji amplitude i faze radio signala vrlo niskih} (VLF, 3 -30 kHz)  {\rrm i/ili niskih frekvencija}, (LF, 30 - 300 kHz){\rrm a koji su posledica naglog  porasta intenziteta }X - {\rrm zraqenja u toku Sunqevog flera, prouqavani su u cilju odredjivanja uslova prostiranja radio signala u} D - {\rrm oblasti jonosfere.} {\rrm U toku naglog i intenzivnog porasta } X {\rrm zraqenja dolazi do porasta gustine elektrona u} D -{\rrm oblasti jonosfere. U Beogradu je instalisana stanica} AWESOME {\rrm juna 2008. godine, koja kontinuirano registruje amplitude i faze radio signala u navedenim frekventnim opsezima, a iste emituju transmiteri rasporedjeni na razliqitim lokacijama u svetu. U ovom radu prikazani su rezultati prouqavanja porem{\cc}aja amplitude i faze na} ICV/20.27 kHz  {\rrm radio signalu koji emituje predajnik u} Isola di Tavolara  {\rrm i na } NSC/45.90 kHz {\rrm radio signalu emitovanog sa transmitera na Siciliji (oba se nalaze u Italiji).} VLF/LF {\rrm radio signali se prostiru kroz talasovod qija je donja  granicna oblast  povr{\ss}ina Zemlje a gornja} D - {\rrm oblast jonosfere. U vremenskom period od juna 2008 do februara 2014. godine izabrano je oko 200 dogadjaja u toku kojih je doslo do naglog porasta intenzuteta} X {\rrm zraqenja. Navedeni {\ss}estogodisnji period obuhvat minimum i maksimum Sunqeve aktivnosti). Naglasavamo da su to bili Sunqevi flerovi klasifikovani kao mali} C {\rrm i umereni} M {\rrm flerovi. Odredjivanje stepena porasta amplitude i faze} VLF/LF {\rrm radio signala u toku Sunqevog flera koristili smo za odredjivanje promene gustine elektrona u funkciji visine u} D - {\rrm oblasti jonosfere, kao i odredjivanje gustine elektrona na zadatoj visini u funkciji vremena pre u toku i posle pojave Sunqevog flera.}

}
\end{multicols}

\end{document}